\documentclass[aps,prl,twocolumn,dvipsnames,superscriptaddress,floatfix,table]{revtex4-2}
\usepackage[T1]{fontenc}
\usepackage[latin9]{inputenc}
\usepackage{color}
\usepackage{ulem}
\usepackage{babel}
\usepackage{amsmath}
\usepackage{amssymb}
\usepackage{graphicx}
\usepackage{esint}
\usepackage{makecell}
\usepackage{multirow}
\usepackage{booktabs}

\usepackage[unicode=trueam,
 bookmarks=false,
 breaklinks=true,pdfborder={0 0 1},backref=false,colorlinks=true]
 {hyperref}
\hypersetup{
 pdfcreator={},pdfproducer={LaTeX with hyperref},linkcolor=blue,anchorcolor=blue,citecolor=blue,filecolor=red,menucolor=red,urlcolor=blue,pdfstartview=FitV,pdfhighlight=/I,pdfpagelayout=TwoColumns,hypertexnames=true}

\makeatletter
\usepackage{lmodern}
\usepackage{mathptmx, newtxtext, newtxmath, xspace}
\usepackage{bbm}

\usepackage[table, dvipsnames]{xcolor}
\definecolor{colorA}{cmyk}{0,0,0,0.05}
\definecolor{colorB}{cmyk}{0.14,0.04,0,0}
\definecolor{colorC}{cmyk}{0.02,0.0799,0,0}
\definecolor{colorD}{cmyk}{0.099,0.14,0,0}
\usepackage{siunitx}
\usepackage{natbib}
\setcitestyle{numbers}

\usepackage{booktabs}
\usepackage{dashrule} 

\makeatother

\begin{document}
\title{Anomalous Hall viscosity 
of altermagnets}
\author{Iksu Jang}
\affiliation{Institute for Theory of Condensed Matter, Karlsruhe Institute of Technology,
Karlsruhe 76131, Germany}
\author{Rui Aquino}
\affiliation{Department of Physics, The Grainger College of Engineering, University of Illinois Urbana-Champaign, Urbana, Illinois 61801, USA}
\affiliation{Anthony J. Leggett Institute for Condensed Matter Theory, The Grainger College of Engineering, University of Illinois Urbana-Champaign, Urbana, Illinois 61801, USA}
\affiliation{ICTP South American Institute for Fundamental Research, S\~ao Paulo, SP, Brazil}
\author{J{\"o}rg Schmalian }
\affiliation{Institute for Theory of Condensed Matter, Karlsruhe Institute of Technology,
Karlsruhe 76131, Germany}
\affiliation{Institute for Quantum Materials and Technologies, Karlsruhe Institute
of Technology, Karlsruhe 76131, Germany}
\author{Rafael M. Fernandes}
\affiliation{Department of Physics, The Grainger College of Engineering, University of Illinois Urbana-Champaign, Urbana, Illinois 61801, USA}
\affiliation{Anthony J. Leggett Institute for Condensed Matter Theory, The Grainger College of Engineering, University of Illinois Urbana-Champaign, Urbana, Illinois 61801, USA}
\date{\today }
\begin{abstract}
We show that the phonon Hall viscosity at zero magnetic field is a natural probe of altermagnetism. First, we demonstrate that the finite elements of the Hall viscosity tensor unambiguously distinguish altermagnets from ferromagnets and conventional antiferromagnets. We then microscopically compute the Hall viscosity in models for $d$-wave and $g$-wave altermagnets, and find a strong sensitivity to electronic spectrum features such as gapped Dirac points and Lifshitz transitions. This sensitivity reflects a strain-space Berry curvature monopole, which contrast to the multipolar character of the standard momentum-space Berry curvature in altermagnets. Since the Hall viscosity can be probed experimentally through magneto-acoustic measurements, it provides 
a compelling method to probe the broken symmetries and topology of insulating altermagnets.
\end{abstract}
\maketitle
Altermagnets are compensated magnets that are invariant under a combination of time reversal and a crystalline operation that involves rotations, such as proper rotations, mirror reflections, glide reflections, and screw rotations~\cite{Smejkal2022,Smejkal2022b}. This symmetry endows altermagnets with distinctive  $d$-wave, $g$-wave, or $i$-wave characters, manifested both in the momentum dependence of the spin-splitting of the electronic bands and in the real-space spin density~\cite{Jungwirth2025altermagnetism,Jungwirth2026symmetry}. However, the experimental challenges in directly probing momentum-space spin-splitting motivates the search for global response functions that not only encode the symmetry of altermagnets, but also their unique topological properties~\cite{Cano2024,Antonenko2024,Agterberg2024,Knolle2024,SchnyderTopology,Farajollahpour2025,Takahashi2025}. 

The anomalous Hall conductivity, a primary probe of non-trivial topology~\cite{Nagaosa2010,vsmejkal2022anomalous}, vanishes in pure altermagnets due to the multipolar character of their momentum-space Berry curvature~\cite{Takahashi2025}. In the presence of spin-orbit coupling (SOC), it can become non-zero either for specific moment directions ~\cite{vsmejkal2020crystal,Kluczyk2023,Attias2024,Roig2025,Osin2025} or when external uniaxial strain is applied~\cite{Takahashi2025}. Nevertheless, most altermagnetic candidates are insulators~\cite{Smejkal2022b,Facio2023,Sodequist2024,Haule2024}, which limits the use of the Hall conductivity as a probe of altermagnetism. Lattice responses, on the other hand, can be measured in both metals and insulators. It is thus valuable to elucidate the lattice responses of altermagnets, particularly given their unique magneto-elastic properties, 
of which piezomagnetism is the posterchild~\cite{ma2021multifunctional,Bhowal2024,Steward2023,fernandes2024topological,aoyama2024piezomagnetic,mcclarty2024landau,vandenBrink2024,Khodas2024,Chakraborty2025,Jiang2025strain,Schiff2025,Hu2025catalog,Karetta2025,Radhakrishnan2026,Khodas2026,Ohlendorf2026}.

\begin{figure}
    \centering
    \includegraphics[width=0.75\linewidth]{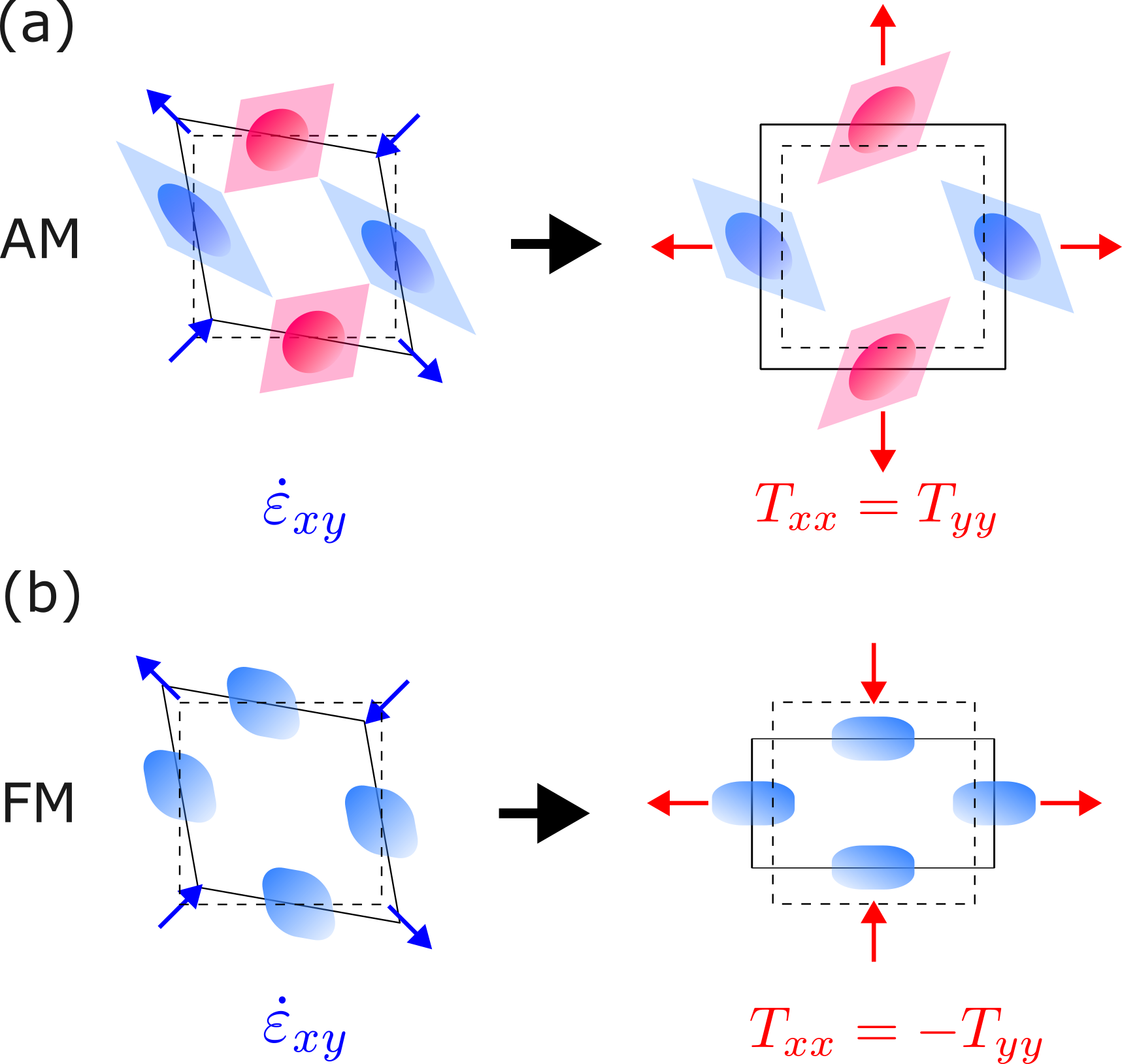}
    \caption{Non-dissipative stress components $T_{xx}$ and $T_{yy}$  generated by a dynamic shear strain $\dot{\varepsilon}_{xy}$ due to the anomalous Hall viscosity in a tetragonal system. Red and blue refer to spin-up and spin-down densities. The stress symmetry is determined by the type of magnetic state:   in a $d$-wave altermagnetic (AM) state, a symmetry-preserving static stress $T_{xx}=T_{yy}$ is generated (a), while in a ferromagnetic (FM) state, tetragonal-symmetry-breaking stress $T_{xx}=-T_{yy}$ appears (b). }
    \label{fig:schematic_tet}
\end{figure}

Here, we show that the Hall viscosity, which is measured through sound waves rather than static strain, provides a natural probe of altermagnetic order and of the underlying topology of the band structure.
Also known as phonon Hall viscosity, it is the antisymmetric, non-dissipative part of the viscosity tensor that relates time-dependent strain to transverse stress (Fig.~\ref{fig:schematic_tet}). While previous works focused on the Hall viscosity of magnetic insulators and quantum Hall systems in an external magnetic field~\cite{Avron1995,Barkeshli2012,Bradlyn2012,Qin2012,Ryu2015,Nagaosa2019,KimBomSoo2020,Ye2021,Scheurer2021,Flebus2023,Kim2024}, our focus is on the zero-field (i.e., anomalous) response. In contrast to antiferromagnets, which cannot have an anomalous Hall viscosity, we find that ferromagnets and altermagnets display finite but distinct Hall viscosity tensor elements at zero field. To elucidate the microscopic mechanism, we calculate the anomalous Hall viscosity in representative minimal models for tetragonal $d$-wave and hexagonal $g$-wave altermagnetism. In both cases, the Hall viscosity reflects a net strain-space Berry curvature monopole, in sharp contrast to the multipolar character of the momentum-space Berry curvature that governs the anomalous Hall conductivity.

This connection with the strain-space Berry curvature makes the Hall viscosity highly sensitivity to special features of the altermagnetic electronic states. In $d$-wave altermagnets, we find it to be dominated by the spin-polarized Dirac points that are gapped by the SOC~\cite{Antonenko2024}. This allows us to interpret the Hall viscosity as the ``Hall conductivity'' due to emergent  spin-dependent electromagnetic gauge fields generated by dynamic strain. In $g$-wave altermagnets, we show that the strain-induced modification of the SOC is the main feature  responsible for a finite anomalous Hall viscosity, which is strongly impacted by the saddle-points of the electronic bands. Importantly, while SOC is essential for a non-zero Hall viscosity, weak SOC does not necessarily imply a small Hall viscosity. 
In fact, the magnitude of our calculated Hall viscosity is comparable to what was recently reported  for  $\alpha$-RuCl$_3$ using the acoustic Faraday effect~\cite{Shragai2026}.

\paragraph*{Symmetry analysis:} At low frequencies, dynamical stresses $T_{ij}\left(t\right)$ in a many-body system are related to time-dependent strain $\varepsilon_{kl }\left(t\right)$ via 
\begin{equation}
T_{ij}\left(t\right)=C_{ijkl}\varepsilon_{kl}\left(t\right)+\eta_{ijkl}\partial_t\varepsilon_{kl}\left(t\right).
    \label{eq:stress_strain}
\end{equation}
The first term is the elastic tensor, whereas the second term usually accounts for the fact that deformations performed at a finite rate are dissipative and generate heat. However, the elements $\eta^H_{ijkl}$ of the viscosity tensor $\eta_{ijkl}$ that are antisymmetric under $ij\longleftrightarrow kl$ are non-dissipative. This Hall viscosity tensor is odd under time reversal and even under inversion, giving rise to the effective elastic
action 
\begin{equation}
S_{{\rm eff}}=\frac{1}{2}\int_{x,t}\eta^H_{ijkl}\varepsilon_{ij}\partial_{t}\varepsilon_{kl},
\label{eq:action_1}
\end{equation}
with  $\int_{x,t}$ denoting integration over space and time. This elastic (phonon) Hall viscosity should not to be confused with the viscoelastic Hall response of an electronic fluid in the hydrodynamic regime~\cite{Mueller2009,Bradlyn2012,Link2016,Sukhachov2025}. Since a finite Hall viscosity requires broken time-reversal symmetry (TRS), it has often been investigated as a response to an external magnetic field~\cite{Nagaosa2019,Scheurer2021,Ye2021,Flebus2023}. For instance, in a 2D isotropic system subjected to a magnetic field, the only allowed non-zero element is $\eta^H_{xxxy}=-\eta^H_{yyxy}$, corresponding to the stress-strain relationship shown in Fig.~\ref{fig:schematic_tet}(b). By symmetry, a ferromagnet must display the same non-zero element, but at zero applied field, corresponding to an anomalous Hall viscosity. However, there are other
ordered phases that break TRS, such as antiferromagnetic and altermagnetic orders. While the combination of time-reversal and translational (or inversion) symmetries enforce $\eta^H_{ij kl}=0$ in antiferromagnets, the symmetries of altermagnetism are generally compatible with finite Hall viscosity tensor elements at zero fields.

To show this, we employ group theory to decompose the 15-dimensional reducible representation consisting of the non-zero elements of  $\eta^H_{ij kl}$  into irreducible representations (irreps) of the point group describing the paramagnetic phase. This task is greatly simplified by expressing 
 $\eta^H_{ijkl}$ in terms of the so-called Jahn symbols~\cite{Jahn1949} as $a\left\{ \left[V^{2}\right]\left[V^{2}\right]\right\}$~\footnote{$V$ stands, as usual, for a polar vector, square brackets imply a symmetric
combination, curly brackets imply an anti-symmetric combination, and
$a$ denotes time reversal.}. 
For example, in a tetragonal system with point group $D_{4h}$, we find:
\begin{equation}
a\left\{ \left[V^{2}\right]\left[V^{2}\right]\right\} \rightarrow A_{1g}^{-}\oplus2 A_{2g}^{-}\oplus2B_{1g}^{-}\oplus2B_{2g}^{-}\oplus4E_{g}^{-},
\label{eq:Jahn}
\end{equation}
where the minus superscript denotes a TRS-odd irrep. These irreps correspond to different even-parity $\boldsymbol{Q}=0$ magnetic orders in the presence of SOC~\cite{fernandes2024topological}. $A_{2g}^{-}$ and $E_{g}^{-}$ transform like the out-of-plane and in-plane magnetization, respectively, thus describing either ferromagnetic or mixed altermagnetic order parameters (i.e., altermagnets with weak ferromagnetism due to SOC). In contrast, $A_{1g}^{-}$, $B_{1g}^{-}$, and $B_{2g}^{-}$ describe pure altermagnetic order parameters (i.e., with a symmetry-enforced zero magnetization) with, respectively, $g$-wave, $d_{xy}$-wave, and $d_{x^2-y^2}$-wave spin-splitting symmetry.  Thus, Eq.~\eqref{eq:Jahn} implies that each type of magnetic order parameter is associated with a unique combination of non-zero anomalous Hall viscosity tensor elements, demonstrating the suitability of $\eta^H_{ij kl}$ to distinguish between magnetic orders. Consider, for instance, the cases of a $d_{x^2-y^2}$ altermagnet ($B_{2g}^-$) and a ferromagnet with out-of-plane magnetization ($A_{2g}^-$). While the same anomalous Hall viscosity elements are non-zero in both cases, they have different relative signs: $\eta^H_{xxxy}=\eta^H_{yyxy}\neq 0$ for the altermagnet and $\eta^H_{xxxy}=-\eta^H_{yyxy}\neq 0$ for the ferromagnet, resulting in different stresses generated by the same dynamic shear strain, see Fig.~\ref{fig:schematic_tet}.
The End Matter shows the symmetry-allowed $\eta^H_{ij kl}$ for altermagnets in various point groups.

\paragraph*{Microscopic expression:} To investigate the microscopic origin of the anomalous Hall viscosity in altermagnets, we consider a system described by Bloch states $\left|u_{\boldsymbol{k},b}\right\rangle $ with crystal
momentum $\boldsymbol{k}$ and band index $b$. The full Hamiltonian is ${\cal H}_{\boldsymbol{k}}={\cal H}_{\boldsymbol{k,}0}+{\cal H}_{\boldsymbol{k,}\varepsilon}$ with a strain-free contribution ${\cal H}_{\boldsymbol{k,}0}=\sum_{b}\xi_{\boldsymbol{k}l}\left|u_{\boldsymbol{k},b}\right\rangle \left\langle u_{\boldsymbol{k},b}\right| $  and a strain-dependent term ${\cal H}_{\boldsymbol{k,}\varepsilon}=\varepsilon_{ij}\sum_{bb'}\gamma_{bb'}^{ij}\left(\boldsymbol{k}\right)\left|u_{\boldsymbol{k},b}\right\rangle \left\langle u_{\boldsymbol{k},b'}\right| $. The coupling matrices  $\gamma_{bb'}^{ij}$
determine the electron-phonon coupling to acoustic lattice vibrations in the band basis~\cite{Takahashi2025}. The stress tensor $T_{ij}$ generated by a dynamic strain that slowly varies in time can then be obtained through the quasi-adiabatic expansion~\cite{Avron1995}:
\begin{equation}
T_{ij}=\left\langle \frac{\delta{\cal H}_{\boldsymbol{k}}}{\delta\varepsilon_{ij}}\right\rangle =\frac{\delta \left\langle {\cal H}_{\boldsymbol{k}}\right\rangle}{\delta\varepsilon_{ij}}+\sum_{kl,b}\Omega_{ijkl}^{\left(b\right)}\left(\boldsymbol{k}\right)\partial_{t}\varepsilon_{kl},
\end{equation}
from which we readily identify the Hall viscosity
\begin{equation}
\eta^H_{ijkl}=\frac{\hslash}{V}\sum_{\boldsymbol{k}b}f\left(\xi_{\boldsymbol{k}b}\right)\Omega_{ijkl}^{\left(b\right)}\left(\boldsymbol{k}\right)\, ,\label{eq:HallViscosityEq}
\end{equation}
and the strain-space Berry curvature
\begin{eqnarray}
\Omega_{ijkl}^{\left(b\right)}\left(\boldsymbol{k}\right)&=&i\left(\left\langle \partial_{\varepsilon_{ij}}u_{\boldsymbol{k},b}\mid\partial_{\varepsilon_{kl}}u_{\boldsymbol{k},b}\right\rangle -\left\langle \partial_{\varepsilon_{kl}}u_{\boldsymbol{k},b}\mid\partial_{\varepsilon_{ij}}u_{\boldsymbol{k},b}\right\rangle \right)\nonumber \\
&=& 2{\rm Im}\sum_{b'\neq b}\frac{\gamma_{b'b}^{ij}\left(\boldsymbol{k}\right)^{*}\gamma_{b'b}^{kl}\left(\boldsymbol{k}\right)}{\left(\xi_{\boldsymbol{k}b}-\xi_{\boldsymbol{k}b'}\right)^{2}},\label{eq:OmegaEq}
\end{eqnarray}
where $V$ is the volume and $f$ is the Fermi-Dirac distribution. In Ref.~\cite{SM}, we derive this result without invoking the quasi-adiabatic approximation. In order for the Hall viscosity in Eq.~\eqref{eq:OmegaEq} to be finite,  ${\cal H}_{\boldsymbol{k,}\varepsilon}$ must not commute with ${\cal H}_{\boldsymbol{k,}0}$, otherwise $\gamma_{b'b}^{ij}$ is diagonal. Moreover, SOC must be present to induce local Berry curvature. To elucidate the microscopic properties that govern $\Omega_{ijkl}^{\left(b\right)}$, we compute it in minimal models for tetragonal $d$-wave and hexagonal $g$-wave altermagnets.

\begin{figure}
    \centering
    \includegraphics[width=\columnwidth]{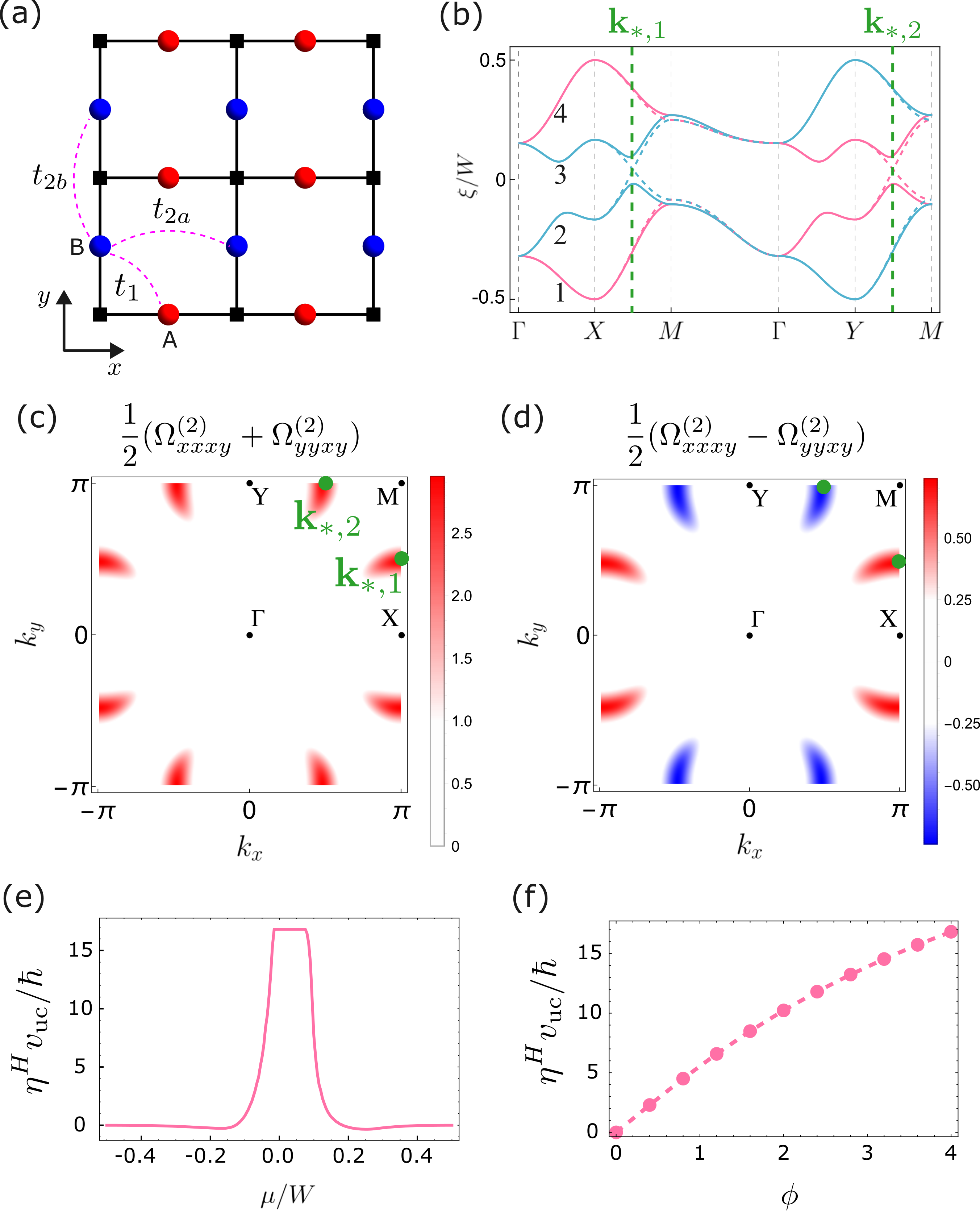}
    \caption{(a) Lieb-lattice model for $d$-wave altermagnetism~\cite{Antonenko2024} and (b) the corresponding band structure with bands labeled by (1,2,3,4). Pink and blue refer to spin up and down states, respectively. The solid (dashed) lines correspond to non-zero (zero) SOC. The Dirac points are located along the green dashed lines. (c) Strain-space Berry curvature elements of the second band, (c) $\frac{1}{2}(\Omega_{xxxy}^{(2)}+\Omega_{yyxy}^{(2)})$ and (d) $\frac{1}{2}(\Omega_{xxxy}^{(2)}-\Omega_{yyxy}^{(2)})$.  (e) Hall viscosity $\eta^H\equiv \frac{1}{2}(\eta^H_{xxxy}+\eta^H_{yyxy})=\eta^H_{xxxy}=\eta^H_{yyxy}$ as a function of the chemical potential and (f) of the altermagnetic order parameter $\phi$ at $\mu=0$. Here, $v_{\mathrm{uc}}$ is the unit cell volume. The values of the tight binding parameters used here are shown in the SM~\cite{SM}.}
    \label{fig:D4hSystem}
\end{figure}

\paragraph*{Tetragonal $d$-wave altermagnets:} The Lieb lattice model of Ref.~\cite{Antonenko2024} (see also~\cite{Sudbo2023,Agterberg2024}) provides a convenient description of a $d$-wave altermagnet on the tetragonal lattice (irrep $B_{2g}^-$), being realized in materials of the families $R$$_2$Mn$_2$Se$_2$O$_3$~\cite{Wei2025,Mazin2025Lieb,Valenti2026},  $A$V$_2$Te$_2$O~\cite{ma2021multifunctional,Zhang2025,Jiang2025,Mazin2026} and Fe$_2$$X$$_2$O ($X$=Cl, Br, I)~\cite{Wang2026}. The model shown in Fig.~\ref{fig:D4hSystem}~(a) has non-magnetic atoms on the sites of a square lattice and magnetic atoms on the bond centers with opposite out-of-plane spins related by a $90^\circ$ crystal rotation. The strain-free Hamiltonian is thus invariant under a combination of a $90^\circ$ rotation and time-reversal~\cite{Antonenko2024,Agterberg2024}
\begin{align}
{\cal H}_{\boldsymbol{k}, 0} = \varepsilon_{0,\boldsymbol{k}} + t_{1,\boldsymbol{k}}\tau_1 + t_{3,\boldsymbol{k}}\tau_3 + \vec{\lambda}_{\boldsymbol{k}} \cdot \vec{\sigma}\, \tau_2 + J \phi \tau_3 \sigma_3 , \label{eq:Lieb}
\end{align}
with sublattice and spin-space Pauli matrices $\tau_i$ and $\sigma_i$, respectively. The first three terms depend on the hopping parameters shown in Fig.~\ref{fig:D4hSystem}~(a) through $\varepsilon_{0,\boldsymbol{k}}=-t_2f_{0,\boldsymbol{k}}$,  $t_{1,\boldsymbol{k}}=-t_1f_{1,\boldsymbol{k}}$, and $t_{3,\boldsymbol{k}}=-t_df_{3,\boldsymbol{k}}$, with $t_2=(t_{2a}+t_{2b})/2$ and $t_{d}=(t_{2a}-t_{2b})/2$, and lattice harmonics $f_{1,\boldsymbol{k}}=4\cos\frac{k_{x}}{2}\cos\frac{k_{y}}{2}$ and $f_{0(3),\boldsymbol{k}}=2\left(\cos k_{x}\pm\cos k_{y}\right)$. The SOC term has the simple form $\vec{\lambda}_{\boldsymbol{k}}= \lambda\sin\frac{k_{x}}{2}\sin\frac{k_{y}}{2} \hat{\boldsymbol{z}} $  and  $\phi$ is the altermagnetic order parameter. The symmetry-allowed strain Hamiltonian contains the three distinct in-plane strain irreps of the tetragonal group (see also~\cite{Takahashi2025,Bell2026})

\begin{equation}
{\cal H}_{\boldsymbol{k,}\varepsilon}=\Big(\varepsilon_{xx}+\varepsilon_{yy}\Big)\gamma^{(A_{1g})}_{\boldsymbol{k}}+\Big(\varepsilon_{xx}-\varepsilon_{yy}\Big)\gamma^{(B_{1g})}_{\boldsymbol{k}}
+  2\varepsilon_{xy}\gamma^{(B_{2g})}_{\boldsymbol{k}}\,
\label{eq:Lieb_strain2}
\end{equation}
with coupling matrices:
\begin{eqnarray}
\gamma^{\left(A_{1g}\right)} _{\boldsymbol{k}}& = & g_{0}^{(A_{1g})}f_{0,\boldsymbol{k}}\tau_{0}+g_{1}^{(A_{1g})}f_{1,\boldsymbol{k}}\tau_{1}+g_{3}^{(A_{1g})}f_{3,\boldsymbol{k}}\tau_{3},\nonumber \\
\gamma^{\left(B_{1g}\right)} _{\boldsymbol{k}} & = & 2g^{(B_{1g})}\tau_{3},\nonumber \\
\gamma^{\left(B_{2g}\right)}_{\boldsymbol{k}} & = & -2g^{(B_{2g})}\sin\frac{k_{x}}{2}\sin\frac{k_{y}}{2}\tau_{1}.
\end{eqnarray}
Here,  $g_{i}^{(\Gamma)}$ are  coupling constants with magnitudes comparable to the hopping parameters~\cite{Takahashi2025}. 
The electronic dispersion of ${\cal H}_{\boldsymbol{k}, 0}$ has four spin-polarized Dirac points at the Brillouin zone boundaries for $|\phi|<\phi_c=|4t_d/J|$, which are gapped by SOC, as shown by the green lines at momenta at $\boldsymbol{k}_{*,1}$ and $\boldsymbol{k}_{*,2}$ in Fig.~\ref{fig:D4hSystem} (b). Being sources of large Berry curvature, these gapped Dirac points determine not only the behavior of the momentum-space Berry curvature quadrupole~\cite{Antonenko2024,Takahashi2025,Mazin2023_FeSe,Radhakrishnan2026}, but also the properties of the strain-space Berry curvature of Eq.~\eqref{eq:HallViscosityEq}. The latter, in turn, gives the non-zero Hall viscosity tensor elements in the altermagnetic phase, which according to our group theory analysis are $\eta^H_{xxxy}=\eta^H_{yyxy}\neq 0$. 

Figs.~\ref{fig:D4hSystem} (c)-(d) show $\frac{1}{2}(\Omega_{xxxy}^{(2)}\pm \Omega_{yyxy}^{(2)})$, respectively,  computed at the second band in Fig.~\ref{fig:D4hSystem} (b). Both show large local Berry curvature values near the gapped Dirac points at $\boldsymbol{k}_{*,1}$ and $\boldsymbol{k}_{*,2}$. While for $\Omega_{xxxy}^{(2)}+ \Omega_{yyxy}^{(2)}$ these values have the same sign near all Dirac points, resulting in a Berry curvature monopole, for $\Omega_{xxxy}^{(2)}- \Omega_{yyxy}^{(2)}$ the Berry curvature changes sign under a $90^\circ$ rotation, resulting in a Berry curvature quadrupole consistent with the quadrupolar structure of the momentum-space Berry curvature $\tilde{\Omega}_{xy}$~\cite{Takahashi2025}. Thus, the strain-space Berry curvature can be uniquely employed to disentangle the intrinsic multipolar structure of the Berry curvature of altermagnets. Using Eq.~\eqref{eq:HallViscosityEq}, these results give $\eta^H_{xxxy}=\eta^H_{yyxy}\neq0$, in agreement with our group-theory analysis. Moreover, as shown in Fig.~\ref{fig:D4hSystem}~(e) and (f), not only is $\eta^H$  proportional to the altermagnetic order parameter $\phi$, but it also is the largest for the chemical potential values for which the system is in the insulating phase. This demonstrates the suitability of exploiting the Hall viscosity to measure altermagnetic order in insulators.

\begin{figure}
    \centering
\includegraphics[width=\linewidth]{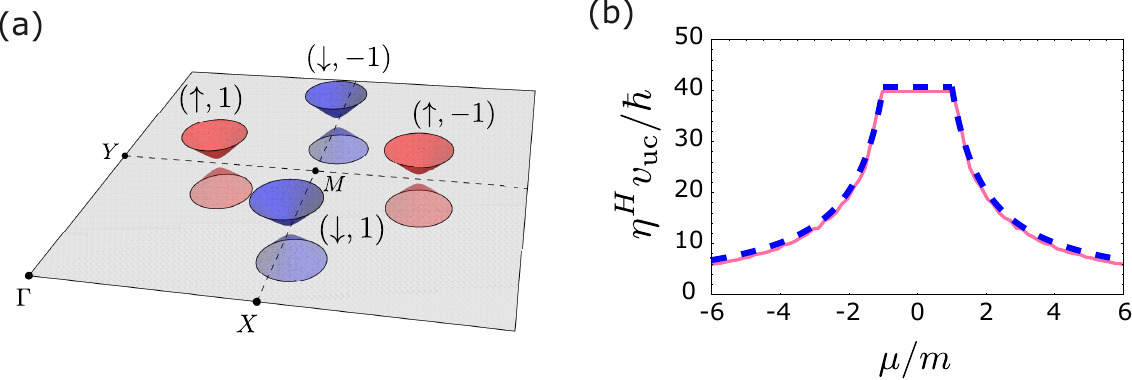}
    \caption{(a) The four spin-polarized Dirac points of the Lieb lattice model are located at the Brillouin zone boundaries when $|\phi| <\phi _c$. They are labeled by their spin ($\sigma=\uparrow,\downarrow$) and valley ($\kappa=\pm1$) quantum numbers ($\sigma$,$\kappa$). (b) Comparison between the Hall viscosity $\eta^H= \eta^H_{xxxy}=\eta^H_{yyxy}$ from the full tight-binding model  (pink solid line) and from the Dirac theory of Eq.~\eqref{eq:HVDirac} (blue dashed line). For the parameters values used here, see the SM~\cite{SM}.}
    \label{fig:Dirac_fig_comp}
\end{figure}

To gain further insights into the topological origin of the Hall viscosity, we expand the Hamiltonian around the Dirac points labeled by valley $\kappa=\pm1 $ and spin $\sigma=\uparrow,\downarrow$, as shown in Fig.~\ref{fig:Dirac_fig_comp}(a). Inclusion of SOC leads to the characteristic Dirac Hamiltonian (see~\cite{SM}):
\begin{equation}
{\cal H}=\sum_{\sigma,\kappa,i=x,y}\left\{v_{i}^{\sigma,\kappa}\left(p_{i}-{\cal A}_{i}^{\sigma,\kappa}\right)\alpha_{i}^{\sigma}+
m^\sigma\beta\right\}
\label{eq:Dirac}
\end{equation}
with Dirac matrices $\alpha_{i}^{\sigma}$ and $\beta$, velocities $v_{i}^{\sigma,\kappa}$, and SOC-generated mass $m^\sigma=\sigma\lambda\sqrt{1-\phi/\phi_c}$. Crucially, strain appears as an electromagnetic gauge field with
\begin{eqnarray}
\varepsilon_{xx}\pm\varepsilon_{yy} & \propto & {\cal A}_{x}^{\uparrow,\kappa}\mp{\cal A}_{y}^{\downarrow,\kappa},\nonumber \\
\varepsilon_{xy} & \propto & {\cal A}_{y}^{\uparrow,\kappa}+{\cal A}_{x}^{\downarrow,\kappa}.
\label{eq:emerg_gauge}
\end{eqnarray}
This is analogous to the emergent magnetic fields in graphene arising from static
but spatially varying strain~\cite{Vozmediano2010,Levy2010}. In our case, because the strain is dynamic and the system is in the altermagnetic phase, the gauge fields correspond to emergent electrical fields, like in Refs.~\cite{vonOppen2009,Cortijo2015,Cortijo2016,Ferreiros2018}, which are however spin-dependent. As a result, the Hall viscosity can be expressed in terms of the Hall conductivity $\sigma^H_{xy}$ (of unit charge $e=1$) of a single Dirac point as:
\begin{equation}
\eta^H_{xxxy}=\eta^H_{yyxy}=C_0 \sigma^H_{xy}=\frac{C_0}{\hslash}\left\{ \begin{array}{ccc}
\frac{{\rm sign}\left(m\right)}{4\pi} & {\rm if} & \mu_{r}^{2}<m^{2}\\
\frac{m}{4\pi\left|\mu_{r}\right|} & {\rm if} & \mu_{r}^{2}>m^{2}
\end{array}\right . \, ,
\label{eq:HVDirac}
\end{equation}
where $\mu_{r}$ is the chemical potential relative to the Dirac point, $m=\lambda\sqrt{1-\phi/\phi_c}$ and $C_0=8\hslash^2 g^{(B_{2g})}g_{3}^{(A_{1g})}/\left(v_{\rm uc}t_{1}t_d\right)$.  As shown in Fig.~\ref{fig:Dirac_fig_comp}(b), this expression is in quantitative agreement with the full tight-binding calculation of the Hall viscosity, at least for small SOC-induced gap~\footnote{Notice that whenever the chemical potential lies within the gap of the Dirac spectrum, the Hall viscosity is independent of both the order parameter $\phi$ and the spin-orbit coupling $\lambda$. However, the limit $\phi \to 0$ cannot be taken, since well-separated Dirac points exist only for sufficiently large values of the order parameter.}.  We emphasize that Eq.~\eqref{eq:HVDirac} does not imply that the system has an anomalous Hall conductivity; instead, it simply expresses the fact that the anomalous Hall viscosity is proportional to the anomalous Hall conductivity of a single  Dirac point, generated by the emergent gauge fields of Eq.~\eqref{eq:emerg_gauge}. Once the contributions of all Dirac points are added, only the anomalous Hall viscosity survives.

It is also instructive to express Eq.~\eqref{eq:action_1}  in terms of the emergent gauge fields:
\begin{equation}
S_{{\rm eff}}=\sigma^{H}_{xy}\sum_{\sigma,\kappa}\sigma\int_{x,t}\epsilon_{ij}{\cal A}_{i}^{\sigma,\kappa}\partial_{t}{\cal A}_{j}^{\sigma,\kappa} \, ,\label{eq:EffCSAction}
\end{equation}
with $\epsilon_{ij}$ the 2D Levi-Civita symbol. This action is noting but the temporal version of two Chern-Simons actions with opposite Hall conductivities for the two spin sectors, ensuring a vanishing Hall conductivity but a finite Hall viscosity.

\begin{figure}
    \centering
    \includegraphics[width=\linewidth]{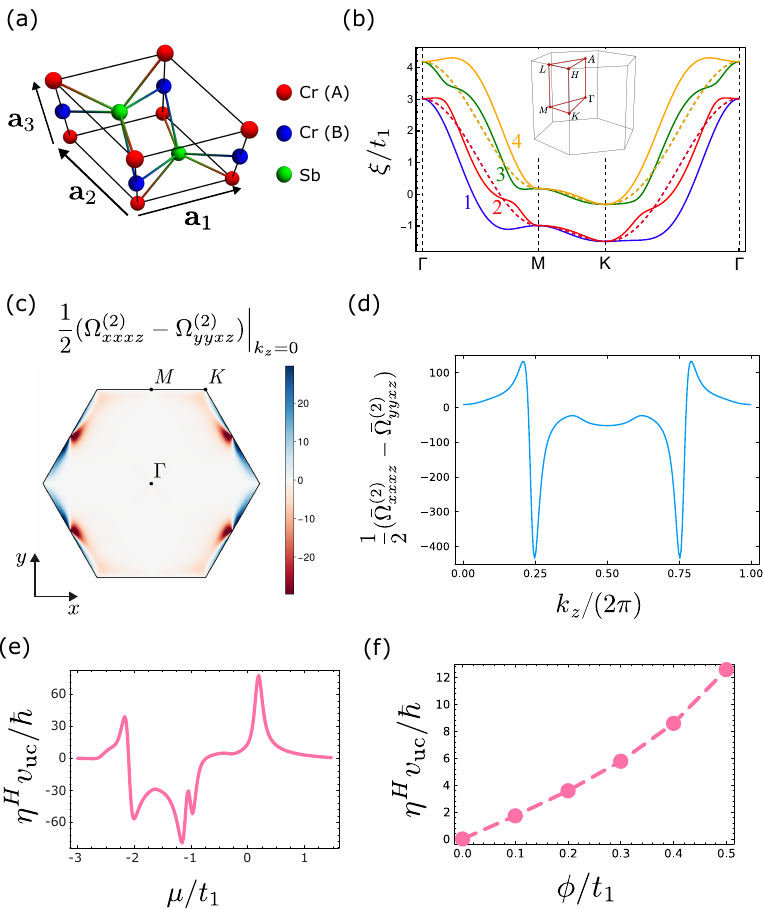}
    \caption{(a) Hexagonal lattice model for $g$-wave altermagnetism. Open and closed symbols refer to atoms away and at the $z=0$ plane, respectively. (b) Band structure of the model with band labels (1,2,3,4); the inset shows the first Brillouin zone. Solid and dashed lines represent the cases with and without SOC, respectively; note that, without SOC, there is no spin-splitting at $k_z=0$. The values of the parameters used here are given in the SM~\cite{SM}. (c) Berry curvature component $(\Omega_{xxxz}^{(4)}-\Omega_{yyxz}^{(4)})/2$. (d) $k_z$-dependence of the Berry curvature $(\bar{\Omega}_{xxxz}^{(4)}-\bar{\Omega}_{yyxz}^{(4)})/2$ integrated over the in-plane momenta. (e) Hall viscosity $\eta^H=\eta^H_{xxxz}=-\eta^H_{yyxz}$ as a function of the chemical potential $\mu$ and (f) of the altermagnetic order parameter $\phi$ for fixed $\mu/t_1=0.5$.}
    \label{fig:D6hSystem}
\end{figure}

\paragraph*{ Hexagonal $g$-wave altermagnets:} The tetragonal $d$-wave altermagnet studied above is essentially a 2D model. In contrast, $g$-wave altermagnetism in hexagonal lattices, as realized in CrSb~\cite{Ding2024large,Yang2024three,Li2024topological,Eaton2026}, MnTe~\cite{mazin2023altermagnetism,krempasky2024altermagnetic,Amin2024nanoscale,Sato2024,Din2025}, and Co$_{1/4}$NbSe$_2$~\cite{Regmi2025,Graham2025,Dale2024,Day2026}, is intrinsically 3D. To investigate their Hall viscosity, we adopt the minimal model of Ref.~\cite{Agterberg2024}, with point group $D_{6h}$.  As illustrated in Fig.~\ref{fig:D6hSystem}~(a), the two magnetic atoms are not on the same plane, but are related by a sixfold screw rotation, i.e., a sixfold rotation followed by a half-translation along the $z$-axis. The Hamiltonian  ${\cal H}_{\boldsymbol{k,}0}$ has the same form of Eq.~\eqref{eq:Lieb}, but the functions depend explicitly on $k_z$ and the SOC term $\vec{\lambda }_{\boldsymbol{k}}$ has in-plane components (see the Supplementary Material~\cite{SM}). 

We focus on moments aligned along the $z$-axis, relevant for CrSb and Co$_{1/4}$NbSe$_2$, in which case the altermagnetic order parameter $\phi$ in Eq.~\eqref{eq:Lieb} transforms as the $B_{1g}^-$ irrep of $D_{6h}$. This gives spin-split nodal planes along $k_z=0$ and $k_y=0$ (and symmetry-related planes), which are partially gapped by SOC, as shown in the electronic dispersion of Fig.~\ref{fig:D6hSystem}~(b). In this case, our group-theory analysis predicts non-zero Hall viscosity tensor components $\eta^H_{xxxz} = -\eta^H_{yyxz} = -\eta^H_{xyyz}$. We therefore consider the coupling matrices $\gamma_{b'b}^{ij}$  in ${\cal H}_{\boldsymbol{k,}\varepsilon}$ associated with the in-plane and out-of-plane shear strains,  $(\varepsilon_{xx}-\varepsilon_{yy},\varepsilon_{xy})$ and $(\varepsilon_{xz},\varepsilon_{yz})$.

Fig.~\ref{fig:D6hSystem}~(c) displays our results for $\frac{1}{2}(\Omega_{xxxz}^{(4)}-\Omega_{yyxz}^{(4)})$ along the $k_z=0$ plane, regularized for clearer presentation (see SM) and calculated at the band $4$ highlighted in Fig.~\ref{fig:D6hSystem}~(b). This Berry curvature component is even with respect to reflections along the $k_x$  and $k_y$  mirrors, and thus averages to a finite value. Fig.~\ref{fig:D6hSystem}~(d) shows the behavior of the in-plane integrated Berry curvature as a function of $k_z$, $\frac{1}{2}(\bar{\Omega}_{xxxz}^{(4)}-\bar{\Omega}_{yyxz}^{(4)})$ where $\bar{\Omega}_{xx(yy)xz}^{(4)}(k_z)=\int_{k_x,k_y}\Omega_{xx(yy)xz}^{(4)}(\boldsymbol{k})$. Its sudden change around $\frac{k_z}{2\pi}\approx 0.25$  originates from a Lifshitz transition of the Fermi surface, which in turn is manifested as sudden changes of the Hall viscosity 
$\eta^H_{xxxz}=-\eta^H_{yyxz}$ 
as a function of the chemical potential $\mu$ in Fig.~\ref{fig:D6hSystem}(e), thus showcasing the sensitivity of the Hall viscosity to the topology of the band structure (a detailed analysis is shown in~\cite{SM}). Importantly, a non-zero $\eta^H$ requires the strain Hamiltonian to also modify the SOC. Without such a strain-induced SOC term, we find a vanishing Hall viscosity. Finally, Fig.~\ref{fig:D6hSystem}(f) directly demonstrates the proportionality between the anomalous Hall viscosity and the altermagnetic order parameter $\phi$.

\paragraph*{Discussion}:  
The calculated Hall viscosity $\eta^H$ in Fig.~\ref{fig:D4hSystem}(e) and Fig.~\ref{fig:D6hSystem}(e) acquires values of the order of $10\,\hslash/v_{{\rm uc}}$ which, taking a typical unit cell volume  $v_{{\rm uc}}=a_{0}^{3}$ with $a_{0}=5$\AA, gives  $\eta^H\sim 8.15\,\mu{\rm Pa}\cdot{\rm s}$. These are comparable to the values recently reported in $\alpha$-RuCl$_3$ under an external magnetic field~\cite{Shragai2026}, measured through the acoustic Faraday effect~\cite{Kitell1958,Crow1971,lee_discovery_1999,Sytcheva2010,Shragai2026}. In a hexagonal system like $\alpha$-RuCl$_3$, this effect arises because the Hall viscosity term $\eta^H_{xxxy}=-\eta^H_{yyxy}$ generated by the magnetic field lifts the degeneracy of the transverse acoustic waves propagating along the $c$-axis, which results in a rotation of the circular polarization of an incident transverse strain wave. In a hexagonal $g$-wave altermagnet like CrSb, however, the zero-field Hall viscosity term $\eta_{xxxz}^{H}=-\eta_{yyxz}^{H}$ does not mix the $c$-axis sound waves. Nevertheless, one can still exploit the near-degeneracy of the transverse modes propagating along a direction that is tilted slightly away from the $c$-axis. Because sound propagation is birefringent along this direction, the linear polarization of an incident transverse wave will generally rotate even in the paramagnetic phase. However, the combination of near-degenerate sound velocities with the form factor of $\eta_{xxxz}^{H}=-\eta_{yyxz}^{H}$ ensures that the altermagnetic contribution to the polarization rotation becomes sizable. A full analysis of magneto-acoustic setups that can be used to probe the Hall viscosity of magnetically ordered states will be presented elsewhere~\cite{Aquino2026}. 

In summary, we have established the anomalous Hall viscosity as a natural bulk geometric response of altermagnets. Our  group-theory analysis enables one to  distinguish between ferromagnets, conventional antiferromagnets, and altermagnets through the Hall viscosity, which in turn can be measured via magneto-acoustic measurements that are similar to the conventional acoustic Faraday effect. Microscopically, we demonstrated that because the Hall viscosity is governed by the strain-space Berry curvature, it becomes highly sensitive to the underlying electronic structure, being strongly enhanced near SOC-gapped Dirac crossings, Lifshitz transitions, and other band-structure singularities. 

These results highlight that the Hall viscosity probes not only symmetry breaking, but also the topological properties of altermagnets. Often, the latter are described in terms of the properties of the momentum-space Berry curvature, which in altermagnets has a multipolar structure. While the magnetic moment direction or strain can distort it to produce a net nonzero momentum-space Berry curvature, the resulting anomalous Hall conductivity can only be measured in metallic or weakly semiconducting materials. In contrast, the strain-space Berry curvature displays a monopole structure even in unstrained pure altermagnets, and the resulting phonon Hall viscosity can be measured also in insulators, which comprise the vast majority of altermagnetic material candidates.

\paragraph*{Acknowledgments:} We are grateful to P. Brouwer, W. J. Meese, D. J. Schultz, J. Sinova, S. Sorn, D. Valentinis, and J. Venderbos for helpful discussions. I.J. and J.S. were supported by the German Research Foundation TRR 288-422213477 ELASTO-Q-MAT,
Project A07. R.A. acknowledges partial support from the Brazilian agencies CNPq through a postdoctoral fellowship (Proc. 201212/2025-0) and from FAPESP (Grant No. 2023/05765-7). R.M.F. acknowledges support from the Research Corporation for Science Advancement through the Cottrell SEED Award CS-SEED-2025-012 and a Mercator Fellowship from the German Research Foundation (DFG) through Grant No. TRR 288,
422213477 Elasto-Q-Mat.

\bibliography{references}

\onecolumngrid
\section*{End Matter}
\appendix
In Table~\ref{tab:class_point_groups} we list the independent non-vanishing  elements of the Hall viscosity tensor $\eta^H_{ijkl} $ generated by different types of ferromagnetic and altermagnetic order in the most symmetric tetragonal, hexagonal, and cubic point groups~\cite{fernandes2024topological}. The presence of SOC is assumed. We used the software Isotropy~\cite{Isotropy} and the Bilbao Crystallographic Server~\cite{Bilbao1} to obtain these results. Note that additional elements follow from the relations $
\eta^H_{ijkl}=\eta^H_{ji kl}=\eta^H_{ijlk}=-\eta^H_{klij}$. 

We consider separately the altermagnetic (AM) and ferromagnetic (FM) order parameters. When the order parameter is multi-dimensional, we show explicitly the order parameter components that are non-zero, e.g. $(1,0,0)$ or $(\sqrt{3},1)$. In this table, the $(x,y,z)$ coordinates refer to the coordinates in the magnetically ordered state, which can be rotated with respect to the coordinates of the paramagnetic phase depending on which order parameter components condense~\cite{Bilbao1}. Note that, for the cubic point group, the $i$-wave AM order parameter that transforms as the $A_{1g}^-$ irrep does not have a non-zero Hall viscosity tensor element. 
\begin{table*}[h]
\centering
\begin{tabular}{lllll}
\hline\hline
Point Group & AM irrep. & Hall viscosity tensor & FM irrep. & Hall viscosity tensor \\
\midrule
\textbf{tet. $D_{4h}$} ($4/mmm$) 
& $A_{1g}^{-}$ ($g$-wave) & $\eta_{xxzz}^{H} = \eta_{yyzz}^{H}$ 
& $A_{2g}^{-}$ & $\eta_{xxxy}^{H} = -\eta_{yyxy}^{H}$, $\eta_{xzyz}^{H}$ \\
& $B_{1g}^{-}$ ($d$-wave) & $\eta_{xxzz}^{H} = -\eta_{yyzz}^{H}$, $\eta_{xxyy}^{H}$ 
& $E_{g}^{-}$ $(1,0)$ & $\eta_{xxxy}^{H}$, $\eta_{yyxy}^{H}$, $\eta_{xzyz}^{H}$, $\eta_{xyzz}$ \\
& $B_{2g}^{-}$ ($d$-wave) & $\eta_{xxxy}^{H} = \eta_{yyxy}^{H}$, $\eta_{xyzz}^{H}$ 
& $E_{g}^{-}$ $(1,1)$ & $\eta_{xxyz}^{H}$, $\eta_{yyyz}^{H}$, $\eta_{zzyz}^{H}$, $\eta_{xyxz}^{H}$ \\
\midrule
\textbf{hex. $D_{6h}$} ($6/mmm$) 
& $A_{1g}^{-}$ ($i$-wave) & $\eta_{xxzz}^{H} = \eta_{yyzz}^{H}$ 
& $A_{2g}^{-}$ & $\eta_{xxxy}^{H} = -\eta_{yyxy}^{H}$, $\eta_{xzyz}^{H}$ \\
& $B_{1g}^{-}$ ($g$-wave) & $\eta_{xxxz}^{H} = -\eta_{yyxz}^{H} = -\eta_{xyyz}^{H}$ 
& $E_{1g}^{-}$ $(1,0)$ & $\eta_{xxyz}^{H}$, $\eta_{yyyz}^{H}$, $\eta_{zzyz}^{H}$, $\eta_{xyxz}^{H}$ \\
& $B_{2g}^{-}$ ($g$-wave) & $\eta_{xxyz}^{H} = -\eta_{yyyz}^{H} = \eta_{xyxz}^{H}$ 
& $E_{1g}^{-}$ $(\sqrt{3},1)$ & $\eta_{xxyz}^{H}$, $\eta_{yyyz}^{H}$, $\eta_{zzyz}^{H}$, $\eta_{xyxz}^{H}$ \\
& $E_{2g}^{-}$ $(1,0)$ ($d$-wave) & $\eta_{xxyy}^{H}$, $\eta_{xxzz}^{H}$, $\eta_{yyzz}^{H}$ 
& & \\
& $E_{2g}^{-}$ $(\sqrt{3},1)$ & $\eta_{xxxy}^{H}$, $\eta_{yyxy}^{H}$, $\eta_{zzxy}^{H}$, $\eta_{xzyz}^{H}$ 
& & \\
\midrule
\textbf{cub. $O_{h}$} ($m\bar{m}m$) 
& $A_{2g}^{-}$ ($d$-wave) & $\eta_{xxyy}^{H} = \eta_{yyzz}^{H} = \eta_{zzxx}^{H}$ 
& $T_{1g}^{-}$ $(1,0,0)$ & $\eta_{xxxy}^{H} = -\eta_{yyxy}^{H}$, $\eta_{xzyz}^{H}$ \\
& $E_{g}^{-}$ $(1,0)$ ($d$-wave) & $\eta_{xxzz}^{H} = \eta_{yyzz}^{H}$ 
& $T_{1g}^{-}$ $(1,1,1)$ & $\eta_{xxxy}^{H} = -\eta_{yyxy}^{H}$, $\eta_{xzyz}^{H}$, \\
& $E_{g}^{-}$ $(\sqrt{3},1)$ & $\eta_{xxzz}^{H} = -\eta_{yyzz}^{H}$, $\eta_{xxyy}^{H}$ 
& & $\eta_{xxxz}^{H} = -\eta_{yyxz}^{H} = \eta_{yzxy}^{H}$ \\
& $T_{2g}^{-}$ $(1,0,0)$ ($d$-wave) & $\eta_{xxxy}^{H} = \eta_{yyxy}^{H}$, $\eta_{xyzz}^{H}$ 
& & \\
& $T_{2g}^{-}$ $(1,1,1)$ & $\eta_{xxzz}^{H} = \eta_{yyzz}^{H}$, $\eta_{xyxz}^{H} = \eta_{xxyz}^{H} = -\eta_{yyyz}^{H}$ 
& & \\
\hline\hline
\end{tabular}
\caption{Non-zero elements of the Hall viscosity tensor $\eta_{ijkl}^{H}$ in point groups $D_{4h}$, $D_{6h}$, and $O_{h}$ for altermagnetic (AM) and ferromagnetic (FM) orders.}
\label{tab:class_point_groups}

\end{table*}

\end{document}


\title{Supplementary Material: Anomalous Hall viscosity of altermagnets}

\author{Iksu Jang}
\affiliation{Institute for Theory of Condensed Matter, Karlsruhe Institute of Technology,
Karlsruhe 76131, Germany}
\author{Rui Aquino}
\affiliation{Department of Physics, The Grainger College of Engineering, University of Illinois Urbana-Champaign, Urbana, Illinois 61801, USA}
\affiliation{Anthony J. Leggett Institute for Condensed Matter Theory, The Grainger College of Engineering, University of Illinois Urbana-Champaign, Urbana, Illinois 61801, USA}
\affiliation{ICTP South American Institute for Fundamental Research, S\~ao Paulo, SP, Brazil}
\author{J{\"o}rg Schmalian }
\affiliation{Institute for Theory of Condensed Matter, Karlsruhe Institute of Technology,
Karlsruhe 76131, Germany}
\affiliation{Institute for Quantum Materials and Technologies, Karlsruhe Institute
of Technology, Karlsruhe 76131, Germany}
\author{Rafael M. Fernandes}
\affiliation{Department of Physics, The Grainger College of Engineering, University of Illinois Urbana-Champaign, Urbana, Illinois 61801, USA}
\affiliation{Anthony J. Leggett Institute for Condensed Matter Theory, The Grainger College of Engineering, University of Illinois Urbana-Champaign, Urbana, Illinois 61801, USA}

\date{\today}

\maketitle

\onecolumngrid
\tableofcontents

\section{Hall viscosity beyond the adiabatic approximation:}
In the main text we used the adiabatic approximation to determine the Hall viscosity, an approach that is well defined for gapped systems, but may be problematic in the presence of gapless excitations. To address this issue we rederive Eq.~(5) without relying on the adiabatic approximation. 
We consider the following  strain-dependent action
\begin{align}
S[\Psi^\dagger,\Psi,\varepsilon]&=\int_{\tau,\mathbf{k}} \Big[\Psi_{\mathbf{k}}^\dagger(\tau)\partial_\tau \Psi_{\mathbf{k}}(\tau)+\Psi_{\mathbf{k}}^\dagger(\tau)\mathcal{H}_{0,\mathbf{k}}\Psi_{\mathbf{k}}(\tau)+\varepsilon_{\alpha\beta}(\tau) \Psi_{\mathbf{k}}^\dagger (\tau)\mathcal{H}_{\varepsilon,\mathbf{k}}^{\alpha\beta}\Psi_{\mathbf{k}}(\tau)\Big].
\end{align}
By integrating out the fermion fields, we obtain the effective action in terms of the strain field $\epsilon$.
\begin{align}
    S_{\rm eff}&=\frac{1}{2}\sum_{i\Omega} \varepsilon_{\alpha\beta}(i\Omega)\varepsilon_{\gamma\delta}(-i\Omega) \Pi^{\alpha\beta,\gamma\delta}(i\Omega),\nonumber \\
    \Pi^{\alpha\beta,\gamma\delta}(i\Omega)&=T\sum_{\mathbf{k},i\omega}{\rm tr}\Big[\mathcal{H}_{\varepsilon,\mathbf{k}}^{\alpha\beta}G_{\mathbf{k},i\omega}\mathcal{H}_{\varepsilon,\mathbf{k}}^{\gamma\delta}G_{\mathbf{k},i\omega+i\Omega}\Big],
\end{align}
where $G_{\mathbf{k},i\omega}=[i\omega-\mathcal{H}_{0,\mathbf{k}}]^{-1}$.
Performing the  Matsubara sum and the analytic continuation $i\Omega\rightarrow \omega+i\gamma$, we obtain
\begin{align}
\Pi^{\alpha\beta,\gamma\delta}(\omega)
&=\sum_{\mathbf{k}, n\neq m} n_F(\xi_{\mathbf{k}n}) \Big[\frac{\langle u_{\mathbf{k},m}|\mathcal{H}_{\epsilon,\mathbf{k}}^{\alpha\beta}|u_{\mathbf{k},n}\rangle\langle  u_{\mathbf{k},n} | \mathcal{H}_{\epsilon,\mathbf{k}}^{\gamma\delta}|u_{\mathbf{k},m}\rangle}{\xi_{\mathbf{k}n}-\xi_{\mathbf{k}m}+\omega+i\eta}-(n\leftrightarrow m)\Big].
\end{align}
Here, $|u_{\mathbf{k},n}\rangle$ and $\xi_{\mathbf{k}n}$ are the $n$-th eigenvector and eigenvalue of $\mathcal{H}_{0,\mathbf{k}}$, respectively. This serves as finite $\omega$ contribution to the elastic tensor. To first order in $\omega$, we obtain 
\begin{align}
    S_{\rm eff}=-\frac{1}{2}\int \frac{d\omega}{2\pi} i\omega \epsilon_{\alpha\beta}(\omega)\epsilon_{\gamma\delta}(-\omega)\eta^{\alpha\beta,\gamma\delta}
\end{align}
from which we readily identify the Hall viscosity
\begin{align}
    \eta^{\alpha\beta,\gamma\delta}&=\sum_{\mathbf{k}}\sum_{n\neq m} \frac{2n_F(\xi_{\mathbf{k}n})}{(\xi_{\mathbf{k}m}-\xi_{\mathbf{k}n})^2} {\rm Im}\Big[\langle u_{\mathbf{k},m}|\mathcal{H}_{\epsilon,\mathbf{k}}^{\alpha\beta}|u_{\mathbf{k},n}\rangle\langle  u_{\mathbf{k},n} | \mathcal{H}_{\epsilon,\mathbf{k}}^{\gamma\delta}|u_{\mathbf{k},m}\rangle\Big]
\end{align}
in agreement with Eqs.~(2) and (5).

\section{Tetragonal $d$-wave altermagnetic model}

\subsection{Numerical values of TB parameters}
For the results presented in Fig. 2 in the main text, we used the following numerical values of parameters
\begin{align}
      t_2=t_1/2,\quad t_d=2t_1, \quad \lambda=2t_1,\quad J=t_1,\quad \phi=\phi_c/2\\  g_{1}^{(A_1)}=g_{1}^{(B_2)}=\alpha t_1,\quad g_{3}^{(A_1)}=\alpha t_d,\quad g_{0}^{(A_1)}=\alpha t_2
\end{align}
with $t_1=1$ and $\alpha =8$.

For Fig.~3 in main text, comparing results obtained from the Dirac theory and the tight binding model, we use following numerical values of parameters:
\begin{align}
    t_2=0,\quad t_d=0.705t_1,\quad \lambda=0.025t_1,\quad J=t_1/2,\quad \phi=\phi_c/2, 
\end{align}
which gives $v_1=v_3=1.41 t_1,\quad m=0.175t_1$, with 
\begin{align}
    \quad g_1^{(A_{1})}=g_1^{(B_2)}=\alpha t_1,\quad g_3^{(A_1)}=\alpha t_d,\quad g_0^{(A_1)}=0,
\end{align}
where $\alpha=8$.

\subsection{Dirac theory and emergent gauge fields}
Here, we expand the Lieb-lattice Hamiltonian Eqs.~(7) and (8) in
the vicinity of the Dirac points. In the absence of spin orbit coupling,
i.e. for $\lambda=0$, there are four Dirac nodes at the boundary of the Brillouin zone. To make the analysis easier, we consider the Brillouin zone defined by $k_x,k_y\in(0,2\pi)$ as shown in Fig.~3(a). Let us assume without restriction that $t_d,J,\phi>0$.
Provided $0<\phi<\phi_{c}=4t_{d}/J$, spin up states have two Dirac
points located at $\left( k_{0},\pi\right)$ and $\left( 2\pi-k_{0},\pi\right)$, while spin down
states have Dirac points at $\left(\pi, k_{0}\right)$ and $\left(\pi, 2\pi-k_{0}\right)$, as depicted in Fig.~3(a), with the momentum scale $k_{0}=\arccos\left(2\phi/\phi_c-1\right)$.
For $J=0$ ($\phi_c=\infty$), $k_{0}=\pi$ and the Dirac points merge at $\left(\pm\pi,\pm\pi\right)$
to become quadratic band touching points. In the vicinity of these
Dirac points and for small spin-orbit coupling, we obtain 
\begin{align}
{\cal H}_{0}^{\uparrow,\kappa}(\mathbf{p})=\kappa v_{1}p_{y}\tau_{1}+\kappa v_{3}p_{x}\tau_{3}+m\tau_{2},
\end{align}
where $\kappa=\pm 1$ stands for the valley indices ($\kappa=+1$: $(k_0,\pi)$, $\kappa=-1$: $(2\pi-k_0,\pi)$) and $\mathbf{p}$ is a momentum vector measured with respect to each valley momentum. The velocities are given as $v_{1}=2t_{1}\sqrt{\phi/\phi_{c}}$ and
$v_{3}=4t_d\sqrt{\phi/\phi_c}\sqrt{1-\phi/\phi_c}$, while  $m=\lambda\sqrt{1-\phi/\phi_c}$ is the mass.
For the down spins we have
\begin{align}
{\cal H}_{0}^{\downarrow,\kappa}(\mathbf{p})=\kappa v_{1}p_{x}\tau_{1}-\kappa v_{3}p_{y}\tau_{3}-m\tau_{2},
\end{align}
where $\kappa=\pm1$ stands for the valley points $(\pi,k_0)$ and $(\pi,2\pi-k_0)$ respectively.

The strain terms at the Dirac points are
\begin{eqnarray}
{\cal H}_{\varepsilon}^{\sigma,
\kappa} & = & 2\Big(\varepsilon_{xx}+\varepsilon_{yy}\Big)\left(\overline{g}_{0}\tau_{0}+\sigma \overline{g}_{3}\tau_{3}\right)-4\varepsilon_{xy}\overline{g}_{1}\tau_{1}\nonumber \\
 & + & 2\Big(\varepsilon_{xx}-\varepsilon_{yy}\Big)\overline{g}'_{3}\tau_{3}
\end{eqnarray}
with $\overline{g}_{0}=2g_{0}^{(A_{1g})}\left(\phi/\phi_{c}-1\right)$,
$\overline{g}_{3}=2g_{3}^{(A_{1g})}\phi/\phi_{c}$, $\overline{g}'_{3}=2g^{(B_{1g})}\left(\phi/\phi_{c}-1\right)$,
and $\overline{g}_{1}=g^{(B_{2g})}\sqrt{1-\phi/\phi_{c}}.$ The
spin dependence of the strain term is a consequence of the fact that
the location of the Dirac points is different for spin-up and spin-down
states. This yields the stress tensors $T^\sigma_{xx+yy}=2\left(\overline{g}_{0}\tau_{0}+\sigma \overline{g}_{3}\tau_{3}\right)$,
$T_{xx-yy}=2\overline{g}_{3}\tau_{3}$, and $T_{xy}=-4\overline{g}_{1}\tau_{1}$. 
The 
Hamiltonian in the presence of stain is then given by Eq.~(10) in the main text, with gauge fields
\begin{eqnarray}
\boldsymbol{{\cal A}}^{\uparrow,\kappa} &=& 2\kappa\left(\frac{\overline{g}'_{3}\left(\varepsilon_{xx}-\varepsilon_{yy}\right)-\overline{g}_{3}\left(\varepsilon_{xx}+\varepsilon_{yy}\right)}{v_{3}},\frac{2\overline{g}_{1}\varepsilon_{xy}}{v_{1}}\right)\nonumber \\
\boldsymbol{{\cal A}}^{\downarrow,\kappa} & =& 2\kappa \left(\frac{2\overline{g}_{1}\varepsilon_{xy}}{v_{1}},\frac{\overline{g}'_{3}\left(\varepsilon_{xx}-\varepsilon_{yy}\right)+\overline{g}_{3}\left(\varepsilon_{xx}+\varepsilon_{yy}\right)}{v_{3}}\right)
\label{eq:EmergentGaugeField}
\end{eqnarray}
and Dirac matrices $\boldsymbol{\alpha}^{\uparrow}=\left(\tau_{3},\tau_{1}\right)$, $\boldsymbol{\alpha}^{\downarrow}=\left(\tau_{1},\tau_{3}\right)$, and 
$\beta=\tau_{2}$, obeying the usual Clifford algebra.
For the velocities, it holds $v^{\uparrow,\kappa}=\kappa \left(v_{3},v_{1}\right)$
while $v^{\downarrow,\kappa}=\kappa \left(v_{1},-v_{3}\right)$. By analyzing the
Hall conductivity of a single gapped Dirac point with respect to these emergent gauge fields, we
obtain the result given in the main text. Using Eq.~\eqref{eq:EmergentGaugeField} and Eq.~(13), we then obtain the following effective action in terms of the strain field:
\begin{align}
    S_{\rm eff}=\frac{8g_1g_3}{t_1t_d}\sigma_{xy}^H\int_{t,x}(\epsilon_{xx}+\epsilon_{yy})\dot{\epsilon}_{xy}
\end{align}
which gives $\eta^{H}_{xx+yy,xy}=\frac{8g_1g_3}{t_1t_d}\sigma_{xy}^H$.

\section{Hexagonal $g$-wave altermagnetic model}
In this section, we provide a detailed description of the hexagonal model for $g$-wave altermagnetism corresponding to space group $P6_3/mmc$ (No. 194), which is relevant for CrSb, following Ref. \cite{Agterberg2024}.

\subsection{Symmetries of CrSb}
\begin{figure}[h]
    \centering
    \includegraphics[width=0.9\linewidth]{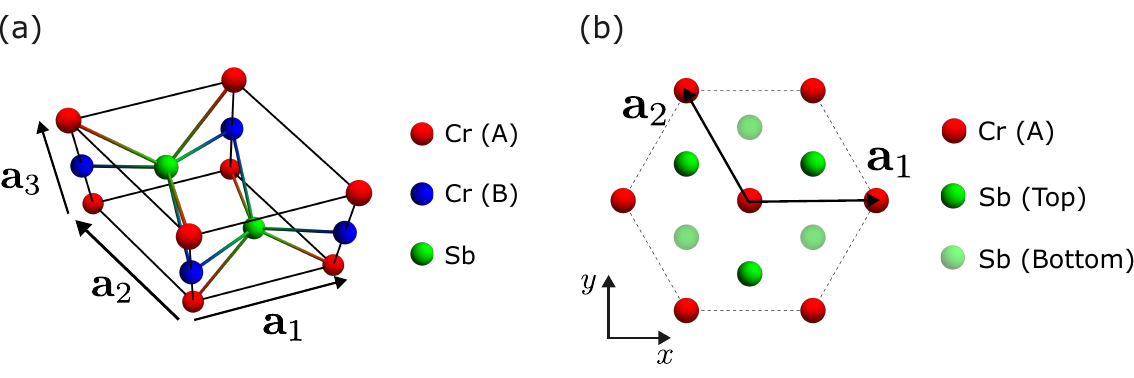}
    \caption{(a) Unit-cell of the CrSb. (b) Top-view of the CrSb lattice.}
    \label{fig:CrSbLatt}
\end{figure}
Fig.~\ref{fig:CrSbLatt} illustrates the unit cell of CrSb, which crystallizes in space group 194 ($P6_3/mmc$), as well as the top-view of the lattice. We define the lattice vectors $\mathbf{a}_1, \mathbf{a}_2$, and $\mathbf{a}_3$ according to the coordinates in Fig.~\ref{fig:CrSbLatt} to construct the space group representations.

The generators of the space group, expressed in Seitz notation $\{R|\mathbf{t}\}$, are:
\begin{equation*}
\mathbb{1}, \quad \{3_{001}|0,0,0\}, \quad \{2_{001}|0,0,1/2\}, \quad \{2_{110}|0,0,0\}, \quad \mathcal{I}
\end{equation*}
where the indices $nml$ for the rotation axes and translation vectors denote the direction $n\mathbf{a}_1 + m\mathbf{a}_2 + l\mathbf{a}_3$. Following Ref.~\cite{Agterberg2024}, we consider the electrons on the two Cr sublattices ($A, B$) as the primary degrees of freedom. Sublattices $A$ and $B$ are located at $(0,0,0)$ and $(0,0,1/2)$, respectively.

The field operators in momentum space are given by:
\begin{equation}
c_{\alpha,\sigma}(\mathbf{k}) = \frac{1}{\sqrt{N}} \sum_{\mathbf{R}} e^{-i\mathbf{k} \cdot (\mathbf{R} + \vec{\eta}_\alpha)} c_{\alpha,\sigma}(\mathbf{R}),
\end{equation}
with sublattice vectors $\vec{\eta}_A = 0$ and $\vec{\eta}_{B} = \frac{1}{2}\mathbf{a}_3$. We thus define the four-component spinor as:
\begin{equation}
\psi(\mathbf{k}) = \begin{pmatrix} c_{A,\uparrow}(\mathbf{k}) & c_{A,\downarrow}(\mathbf{k}) & c_{B,\uparrow}(\mathbf{k}) & c_{B,\downarrow}(\mathbf{k}) \end{pmatrix}^T.
\end{equation}

\begin{table}[h]
\centering
\renewcommand{\arraystretch}{1.4}
\begin{tabular}{|l|l|l|}
\hline
\textbf{Generator} & \textbf{Operator Transformation} & \textbf{Vector Relation} \\ \hline \hline
$\{3_{001}|0,0,0\}$ & $\begin{cases} c_A^\dagger(\mathbf{R}) \to c_A^\dagger(\mathbf{R}') \\ c_B^\dagger(\mathbf{R}) \to c_B^\dagger(\mathbf{R}') \end{cases}$ & $\mathbf{R}'=3_{001}\mathbf{R}, \begin{cases} 3_{001}\mathbf{a}_1=\mathbf{a}_2 \\ 3_{001}\mathbf{a}_2=-\mathbf{a}_1-\mathbf{a}_2 \end{cases}$ \\ \hline
$\{2_{001}|0,0,1/2\}$ & $\begin{cases} c_A^\dagger(\mathbf{R}) \to c_B^\dagger(\mathbf{R}') \\ c_B^\dagger(\mathbf{R}) \to c_A^\dagger(\mathbf{R}'+\mathbf{a}_3) \end{cases}$ & $\mathbf{R}'=2_{001}\mathbf{R}, \begin{cases} 2_{001}\mathbf{a}_1=-\mathbf{a}_1 \\ 2_{001}\mathbf{a}_2=-\mathbf{a}_2 \end{cases}$ \\ \hline
$\{2_{110}|0,0,0\}$ & $\begin{cases} c_A^\dagger(\mathbf{R}) \to c_A^\dagger(\mathbf{R}') \\ c_B^\dagger(\mathbf{R}) \to c_B^\dagger(\mathbf{R}'-\mathbf{a}_3) \end{cases}$ & $\mathbf{R}'=2_{110}\mathbf{R}, \begin{cases} 2_{110}\mathbf{a}_1=\mathbf{a}_2 \\ 2_{110}\mathbf{a}_2=\mathbf{a}_1 \\ 2_{110}\mathbf{a}_3=-\mathbf{a}_3 \end{cases}$ \\ \hline
$\mathcal{I}$ & $\begin{cases} c_A^\dagger(\mathbf{R}) \to c_A^\dagger(\mathbf{R}') \\ c_B^\dagger(\mathbf{R}) \to c_B^\dagger(\mathbf{R}'-\mathbf{a}_3) \end{cases}$ & $\mathbf{R}'=-\mathbf{R}$ \\ \hline
\end{tabular}
\caption{Space-group symmetry operations on the fermionic operators for space group 194, relevant to hexagonal CrSb.}
\label{tab:SymmetryRepresentations}
\end{table}

\begin{table}[h]
\centering
\renewcommand{\arraystretch}{1.5}
\begin{tabular}{|l|l|}
\hline
\textbf{Space group operations (194)} & \textbf{Point group operations ($D_{6h}$)} \\ \hline \hline
$\{6_{001}|0,0,1/2\} = \{2_{001}|0,0,1/2\}\{3_{001}|0,0,0\}$ & $C_6$ \\ \hline
$\{3_{001}|0,0,0\}$ & $C_3$ \\ \hline
$\{2_{001}|0,0,1/2\}$ & $C_2$ \\ \hline
$\{2_{100}|0,0,0\}$, $\{2_{110}|0,0,0\}$, $\{2_{010}|0,0,0\}$ & $C_{2}''$ ($2_{100}$) \\ \hline
$\{2_{\bar{1}10}|0,0,0\}$, $\{2_{210}|0,0,0\}$, $\{2_{120}|0,0,0\}$ & $C_2'$ ($2_{120}$) \\ \hline
$\mathcal{I}$ & $\mathcal{I}$ \\ \hline
\end{tabular}
\caption{Correspondence between the symmetry operations in the space group 194 and the symmetry operations of the point group $D_{6h}$.}
\label{Tab:SpaceGroupPointGroupCorrespondence}
\end{table}

Table~\ref{tab:SymmetryRepresentations} summarizes the action of the symmetry operations of the space group on the fermion operators. In the spinor representation, these operations are:
\begin{align}
\{3_{001}|0,0,0\} \psi(\mathbf{k}) \{3_{001}|0,0,0\}^{-1} &= (\tau_0 \otimes e^{i\frac{\pi}{3}\sigma_z}) \psi(3_{001}\mathbf{k}), \\
\{2_{001}|0,0,1/2\} \psi(\mathbf{k}) \{2_{001}|0,0,1/2\}^{-1} &= e^{i\frac{k_z}{2}} (\tau_x \otimes e^{i\frac{\pi}{2}\sigma_z}) \psi(2_{001}\mathbf{k}), \\
\{2_{110}|0,0,0\} \psi(\mathbf{k}) \{2_{110}|0,0,0\}^{-1} &= (\tau_0 \otimes e^{i\frac{\pi}{2}\sigma_x}) \psi(2_{110}\mathbf{k}), \\
\mathcal{I} \psi(\mathbf{k}) \mathcal{I}^{-1} &= (\tau_0 \otimes \sigma_0) \psi(-\mathbf{k}).
\end{align}

It is convenient to use the correspondence between operations in space group 194 and those of the underlying $D_{6h}$ point group, as shown in Table~\ref{Tab:SpaceGroupPointGroupCorrespondence}. This correspondence is obtained by comparing the character table of each case. Using this correspondence, the fermionic bilinears can be classified into irreducible representations of $D_{6h}$ as follows:
\begin{itemize}
    \item $\psi^\dagger(\mathbf{k})\tau_0\psi(\mathbf{k}): A_{1g}$
    \item $\psi^\dagger(\mathbf{k})\tau_1\psi(\mathbf{k}): A_{1g}$
    \item $\psi^\dagger(\mathbf{k})\tau_2\psi(\mathbf{k}): B^{-}_{2g}$ 
    \item $\psi^\dagger(\mathbf{k})\tau_3\psi(\mathbf{k}): B_{2g}$
\end{itemize}

where the minus superscript indicates a time-reversal odd irrep. 

\subsection{Tight-binding parameters}
Using the spinors introduced above, it is now straightforward to construct the Hamiltonian $H= \sum_{\mathbf{k}} \psi^\dagger(\mathbf{k})\mathcal{H}_{\mathbf{k}}\psi(\mathbf{k})$. Following Ref.~\cite{Agterberg2024}, the strain-free part of the Hamiltonian $\mathcal{H}_{\mathbf{k},0}$ is given by:
\begin{equation}
\mathcal{H}_{\mathbf{k}, 0} = \varepsilon_{0,\mathbf{k}} + t_{1,\mathbf{k}}\tau_1 + t_{3,\mathbf{k}}\tau_3 + \tau_2 \vec{\lambda}_{\mathbf{k}} \cdot \vec{\sigma} + J \phi \tau_3 \sigma_3, 
\label{eq:D6hHam}
\end{equation}
where the hopping terms are:
\begin{align}
\varepsilon_{0,\mathbf{k}} &= t_1 \left( \cos k_x + 2\cos \frac{k_x}{2} \cos \frac{\sqrt{3}k_y}{2} \right) + t_2\cos k_z - \mu, \\
t_{1,\mathbf{k}} &= t_3 \cos \frac{k_z}{2}, \quad t_{3,\mathbf{k}} = t_4 \sin k_z f_y(f_y^2 - 3f_x^2), \\
\lambda_{x,\mathbf{k}} &= \lambda \cos \frac{k_z}{2}(f_x^2 - f_y^2), \quad \lambda_{y,\mathbf{k}} = -2\lambda \cos \frac{k_z}{2} f_x f_y, \quad \lambda_{z,\mathbf{k}} = \lambda_z \sin \frac{k_z}{2} f_x(f_x^2 - 3f_y^2),
\end{align}
and with hexagonal form factors defined as $f_x = \sin k_x + \sin \frac{k_x}{2} \cos \frac{\sqrt{3}k_y}{2}$ and $f_y = \sqrt{3} \cos \frac{k_x}{2} \sin \frac{\sqrt{3}k_y}{2}$. Here, $x$ is the direction parallel to the $\mathbf{a}_1$ in Fig. \ref{fig:CrSbLatt}, corresponding to the crystallographic direction $[100]$. Note that the altermagnetic order parameter $\phi$ transforms as $B_{1g}^- = B_{2g} \otimes A_{2g}^-$, since $\psi^\dagger(\mathbf{k})\tau_3\psi(\mathbf{k})$ transforms as $B_{2g}$, as reflected in the transformation properties of the function $t_{3,\mathbf{k}}$, and $\sigma_3$ transforms as $A_{2g}^-$.

The strain Hamiltonian $\mathcal{H}_{\mathbf{k},\varepsilon}$ consists of two parts corresponding to the shear components that transform as the $E_{1g}$ (out-of-plane shear) and $E_{2g}$ (in-plane shear) irreps of the point group $D_{6h}$:
\begin{align}
    \mathcal{H}_{\mathbf{k}, \varepsilon} &= \mathcal{H}_{\mathbf{k}, \varepsilon}^{(E_{1g})} + \mathcal{H}_{\mathbf{k}, \varepsilon}^{(E_{2g})}, \\
    \mathcal{H}_{\mathbf{k}, \varepsilon}^{(E_{1g})} &= g_{\tau_1}^{(E_{1g})} \left[ \varepsilon_{xz}f_{1,\mathbf{k}}^{(xz)} + \varepsilon_{yz}f_{1,\mathbf{k}}^{(yz)} \right] \tau_1 + g_{\tau_3}^{(E_{1g})} \left[ \varepsilon_{xz}f_{3,\mathbf{k}}^{(xz)} + \varepsilon_{yz}f_{3,\mathbf{k}}^{(yz)} \right] \tau_3, \\
    \mathcal{H}_{\mathbf{k}, \varepsilon}^{(E_{2g})} &= g_{\tau_1}^{(E_{2g})} \left[ (\varepsilon_{xx}-\varepsilon_{yy})f_{1,\mathbf{k}}^{(x^2-y^2)} - 2\varepsilon_{xy}f_{1,\mathbf{k}}^{(xy)} \right] \tau_1 \nonumber \\
    &\quad + g_{\tau_3}^{(E_{2g})} \left[ (\varepsilon_{xx}-\varepsilon_{yy})f_{3,\mathbf{k}}^{(x^2-y^2)} - 2\varepsilon_{xy}f_{3,\mathbf{k}}^{(xy)} \right] \tau_3 \nonumber \\
    &\quad + g_{\text{SOC}}^{(E_{2g})} f_{\text{SOC},\mathbf{k}} \left[ (\varepsilon_{xx}-\varepsilon_{yy})\sigma_x - 2\varepsilon_{xy}\sigma_y \right] \tau_2
\end{align}
where $g_{i}^{(\Gamma)}$ are the coupling constants and the form factors $f_{i,\mathbf{k}}^{(\alpha)}$ are:
\begin{align}
f_{1,\mathbf{k}}^{(xz)} &= -\frac{8}{\sqrt{3}}\sin\frac{k_z}{2}\sin\frac{k_x}{2}\left(2\cos\frac{k_x}{2}+\cos\frac{\sqrt{3}k_y}{2}\right), & f_{3,\mathbf{k}}^{(xz)} &= -4\sin\frac{k_x}{2}\sin\frac{\sqrt{3}k_y}{2}, \\
f_{1,\mathbf{k}}^{(yz)} &= -8\sin\frac{k_z}{2}\cos\frac{k_x}{2}\sin\frac{\sqrt{3}k_y}{2}, & f_{3,\mathbf{k}}^{(yz)} &= \frac{4}{\sqrt{3}}\left(\cos k_x-\cos\frac{k_x}{2}\cos\frac{\sqrt{3}k_y}{2}\right), \\
f_{1,\mathbf{k}}^{(x^2-y^2)} &= \frac{8}{\sqrt{3}}\cos\frac{k_z}{2}\left(\cos k_x-\cos\frac{k_x}{2}\cos\frac{\sqrt{3}k_y}{2}\right), & f_{3,\mathbf{k}}^{(x^2-y^2)} &= -\sqrt{3}\sin k_z \cos\frac{k_x}{2}\sin\frac{\sqrt{3}k_y}{2}, \\
f_{1,\mathbf{k}}^{(xy)} &= 8\cos\frac{k_z}{2}\sin\frac{k_x}{2}\sin\frac{\sqrt{3}k_y}{2}, & f_{3,\mathbf{k}}^{(xy)} &= \sin k_z \left(\sin k_x+\sin\frac{k_x}{2}\cos\frac{\sqrt{3}k_y}{2}\right), \\
f_{\text{SOC},\mathbf{k}} &= \cos \frac{k_z}{2}.
\end{align}

The results shown in the main text were obtained for the following set of tight-binding parameters:
\begin{gather*}
 t_2=0.6 t_1,\quad t_3=0.3 t_1,\quad t_4=0.2 t_1,\quad \lambda=\lambda_z=0.5 t_1,\quad \phi_z=0.5t_1,\\
g_{\tau_1}^{(E_{1g})}=g_{\tau_3}^{(E_{1g})}=g_{\tau_1}^{(E_{2g})}=g_{\tau_3}^{(E_{2g})}=g_{\rm SOC}^{(E_{2g})}=8t_1.
\end{gather*}

For the quantitative analysis of the Hall viscosity it is necessary to relate the typical values of the electron-strain couplings $g=\alpha t_1$ to the hopping elements $t_1$. 
Consider hopping elements between orbitals of angular momentum $\ell$ and $\ell'$ at a distance $r$ and with equilibrium distance $r_0$. They behave like $t(r)\sim t_0 (r/r_0)^{-(\ell +\ell'+1)}\approx t_{0}\left(1+\alpha\frac{r-r_{0}}{r_{0}}\right)$~\cite{Harrison2012,Takahashi2025}.
Then, the dimensionless constant is $\alpha=\ell+\ell'+1$. Hence, $\alpha=1$ describes hopping between two $s$-orbitals, while $\alpha=5$ refers to hopping between two $d$-orbitals.  If the hopping is a consequence of a second order process between two orbitals with angular momentum $\ell$ via an intermediate orbital with angular momentum $\ell'$, we obtain $\alpha=2\left(\ell+\ell'+1\right)$.
Hence hopping between two $d$-orbitals ($\ell=2$) via a $p$-orbital
($\ell'=1$) yield the value used here, $\alpha=8$.

\section{Hall viscosity and $k_z$-dependent Lifshitz transitions}
\begin{figure}
\centering
\includegraphics[width=0.9\linewidth]{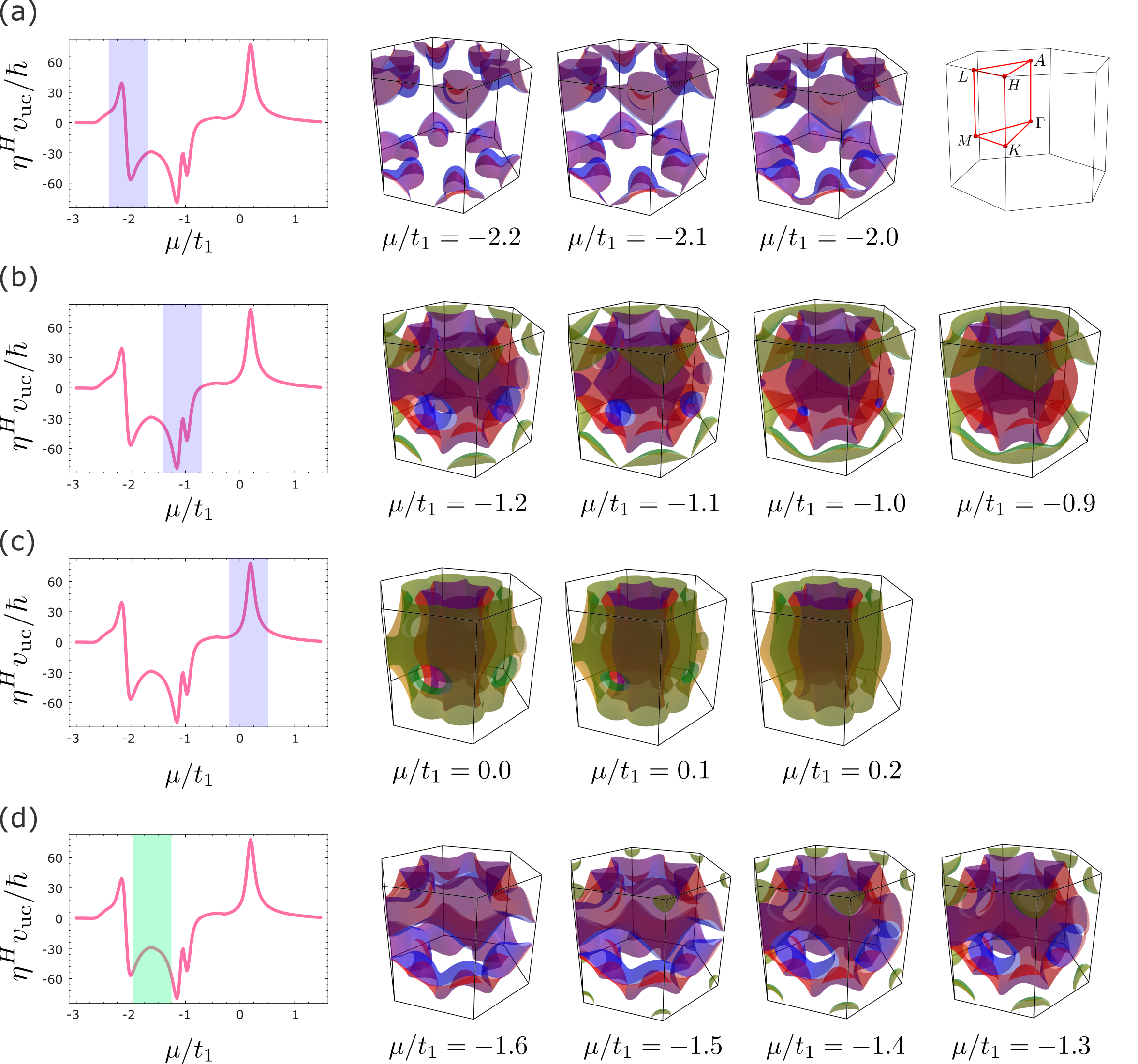}
\caption{Hall viscosity and Lifshitz transitions in the hexagonal $g$-wave altermagnetic model. (a)--(c) Hall viscosity within the chemical potential intervals (a) $-2.3 \lessapprox \mu/W \lessapprox -1.9$, (b) $-1.3 \lessapprox \mu/W \lessapprox -0.8$, and (c) $-0.1 \lessapprox \mu/W \lessapprox 0.3$ (blue-shaded regions), accompanied by the evolution of the Fermi surface. These panels demonstrate that Lifshitz transitions at the $L$ and $M$ high-symmetry points correlate with abrupt changes in the Hall viscosity. (d) In contrast, Lifshitz transitions at the $H$ and $K$ points within the range $-1.7 \lessapprox \mu/W \lessapprox -1.2$ (green-shaded region) do not exhibit a similar behavior.}
\label{fig:Lifshitz_Transition}
\end{figure}

We demonstrate that the abrupt changes in the Hall viscosity of the hexagonal $g$-wave altermagnet are directly linked to Lifshitz transitions. Figs.~\ref{fig:Lifshitz_Transition}(a)--(c) show the changes in the Fermi surface as the chemical potential varies within the narrow range in which the Hall viscosity shows sharp changes. In all cases, we observe Lifshitz transitions from closed to open Fermi surfaces around the high-symmetry $L$ and $M$ points, establishing the connection between Lifshitz transitions and changes in the Hall viscosity. In contrast, when the Lifshitz transition takes place near the $H$ and $K$ points, the Hall viscosity does not display sharp changes, as shown in Fig.~\ref{fig:Lifshitz_Transition}(d). This distinction arises because the strain-dependent Hamiltonian $\mathcal{H}_{\epsilon,\mathbf{k}}$ at these points either vanishes or couples to only a single component of the strain tensor, thereby precluding a Hall response.

\subsection{Details of the Berry curvature calculation}

In the numerical calculation of the Hall viscosity, in order to avoid high symmetry points that usually coincide with time-reversal invariant-momenta, we set our grids as follows:
\begin{equation}
    \mathbf{k}_{ijk} = \mathbf{G}_1 \frac{i - 0.5}{N_1} + \mathbf{G}_2 \frac{j - 0.5}{N_2} + \mathbf{G}_3 \frac{k - 0.5}{N_3}, \quad 
    \begin{cases} 
      i = 1, \dots, N_1 \\ 
      j = 1, \dots, N_2 \\ 
      k = 1, \dots, N_3 
    \end{cases}
\end{equation}
where $\mathbf{G}_1$, $\mathbf{G}_2$, and $\mathbf{G}_3$ are the reciprocal lattice vectors defined as:
\begin{align}
    \mathbf{G}_1=2\pi\left(\hat{x}+\frac{1}{\sqrt{3}}\hat{y}\right),\quad \mathbf{G}_2=\frac{4\pi}{\sqrt{3}}\hat{y},\quad \mathbf{G}_3=2\pi\hat{z}.
\end{align}

We found that the numerical calculation of the Hall viscosity tensor for the hexagonal $g$-wave model converges poorly as a function of grid size. This is because of the change of the projected two-dimensional Fermi surface topology as a function of $k_z$. Near the $k_z$ planes associated with these Lifshitz transitions, the $k_x,\, k_y$-integrated Berry curvature, defined as $\bar{\Omega}(k_z) \equiv \int_{k_x,k_y} \Omega(k_x,k_y,k_z)$, exhibits severe oscillations depending on the grid resolution. Consequently, the Hall viscosity, which is proportional to $\int_{k_z} \bar{\Omega}(k_z)$, converges extremely slowly, failing to stabilize even when the grid dimensions $N_1 \times N_2 \times N_3$ approach $500 \times 500 \times 500$.

To resolve this issue, we introduced the Fermi-Dirac distribution function with temperature scales $T\approx\frac{8t_1}{N_1N_2N_3}$. This thermal broadening acts as a numerical low-pass filter that effectively smooths out the topological transitions across the discrete grid slices. As a result, the Hall viscosity tensor converges significantly faster and remains robust against variations in grid configuration. For the numerical results presented in the main text, we set $N_1 = N_2 = N_3 = 240$, which corresponds to a regularizing temperature scale of $T/t_1 = 5.8 \times 10^{-7}$. 

We note that, although the numerical grid was chosen to avoid all high-symmetry points, certain momentum points still exhibit large local Berry curvature values. This high concentration makes it difficult to clearly identify the underlying symmetry properties of the Berry curvature. To resolve this, we introduce a regularized Berry curvature for Figs.~4(d) and 4(e) in main text, defined as:
\begin{align}
\Omega_{\alpha\beta\gamma\delta}^{\left(l\right),{\rm reg}}\left(\boldsymbol{k}\right)= 2{\rm Im}\sum_{m\neq l}\frac{\gamma_{ml}^{\alpha\beta}\left(\boldsymbol{k}\right)^{*}\gamma_{ml}^{\gamma\delta}\left(\boldsymbol{k}\right)}{\left(\xi_{\boldsymbol{k}l}-\xi_{\boldsymbol{k}m}\right)^{2}+\delta^2}.
\end{align} 

For these figures, we set $\delta=0.5t_1$. Note that this regularized version is strictly used for visual presentation in these specific plots; all calculations for the Hall viscosity tensor presented in the main text were performed using the exact formulation with $\delta=0$.

\bibliography{SM_references}